# Point-by-point inscription of apodized fiber Bragg gratings

**Robert J. Williams,[1,*] Christian Voigtländer,[2] Graham D. Marshall,[1] Andreas Tünnermann,[2] Stefan Nolte,[2] M. J. Steel,[1] and Michael J. Withford[1]**

[1]*Centre for Ultrahigh bandwidth Devices for Optical Systems (CUDOS), MQ Photonics Research Centre, Department of Physics and Astronomy, Macquarie University, New South Wales 2109, Australia*

[2]*Institute of Applied Physics, Friedrich-Schiller-University, Max-Wien-Platz 1, D-07743 Jena, Germany*

*\*Corresponding author: robert.williams@mq.edu.au*

We demonstrate apodized fiber Bragg gratings inscribed with a point-by-point technique. We tailor the grating phase and coupling amplitude through precise control over the longitudinal and transverse position of each laser-inscribed modification. This method of apodization is facilitated by the highly-localized, high-contrast modifications generated by focussed IR femtosecond laser inscription. Our technique provides a simple method for the design and implementation of point-by-point fiber Bragg gratings with complex apodization profiles.

Fiber Bragg gratings (FBGs) inscribed using the point-by-point (PbP) method are becoming increasingly popular for a variety of fiber laser and sensing applications due to the flexibility of the technique [1-4]. Although PbP inscription of FBGs was first demonstrated in 1993 by Malo *et al.* using a UV excimer laser [5], the field received little interest until the advent of IR femtosecond laser materials processing. The use of femtosecond pulses has enabled the inscription of highly-localized micro-void modifications via nonlinear photoionization processes [6]. Coherent Bragg scattering from each of the micro-void modifications in a PbP FBG produces high-quality grating reflection spectra with low loss [7-9]. The femtosecond laser-material interaction, which relies upon nonlinear processes, enables grating inscription into almost any fiber material without requiring photosensitivity [10]. A further advantage of this technique is the high-temperature stability of the gratings due to the structural modification of the glass [11].

The side-band reflection peaks typical of uniform FBG spectra can be problematic for many applications, causing cross-talk in WDM systems, instabilities in Q-switched fiber lasers and linewidth broadening in high power fiber lasers [2,12]. To eliminate these unwanted side-bands it is necessary to fabricate gratings with an apodized profile, where the grating strength varies as a function of length and is typically weaker at both ends of the grating. Various techniques have been developed for fabricating apodized FBGs using holographic inscription, such as phase-mask dithering and the use of phase-masks with varying diffraction efficiency [13,14]; however, these methods are specific to holographic inscription techniques and do not translate directly to PbP inscription. To fabricate apodized PbP gratings one might consider modulating the laser pulse energy during inscription; however, this is likely to be very challenging due to the highly-nonlinear laser-material interaction processes.

In this letter we report apodized fibre Bragg gratings inscribed using a point-by-point technique. Our apodization approach is to tailor the local coupling amplitude of the gratings through precise control over the transverse position of each laser-inscribed refractive index modification (RIM), thereby varying the overlap of the RIM with the core mode. We achieve further flexibility in the apodization profile by introducing discrete phase-shifts in the grating. This technique may in principle be used to achieve apodization profiles of arbitrary complexity. We demonstrate the versatility of our technique by fabricating gratings with Gaussian and sinc $(\text{sinc}(x)=(\sin x)/x)$ apodization profiles. We chose to demonstrate gratings with these apodization profiles as Gaussian-apodized gratings are standard amongst traditional UV-inscribed FBGs and are intrinsically simple to fabricate, while sinc-apodized gratings yield spectra characterized by steep band-edges and a relatively flat-top peak [15], which is ideal for a variety of applications.

The FBG fabrication technique is described in detail in [8]. We stripped the polymer jacket from Corning SMF-28e optical fibre and threaded the fiber through a glass ferrule that was mounted on a piezo-controlled 3-axis flexure translation stage and positioned in front of a 0.8 N.A., 20× oil immersion objective lens. This setup incorporating the glass ferrule is critical to the success of our technique as it provides direct, sub-micron control over the position of the fiber with respect to the focal point of the objective. A high-precision air-bearing stage was used to draw the fiber through the ferrule at a constant velocity while femtosecond laser pulses were focussed into the fiber core at a constant repetition rate. We emphasize that the pulse energy of the femtosecond laser was constant throughout each grating inscription. The laser pulses of <120 fs duration were generated by a regeneratively-amplified Ti:sapphire femtosecond laser operating at 800 nm and a 500 Hz repetition rate. We fabricated second- and third-order gratings with target wavelengths $\lambda_B$ in the range 1520 – 1570 nm (corresponding to periods of approximately 1.1 μm and 1.6 μm respectively). Second- and third-order gratings were used to enable high coupling amplitudes whilst eliminating overlap of the RIMs. The gratings were analyzed in reflection and transmission using a high-resolution (3 pm) and high-sensitivity (>50 dB SNR) C-band swept wavelength system (JDSU 15100).

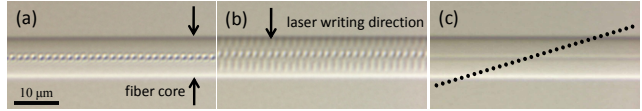

Fig. 1. Transmission differential interference micrographs of a PbP FBG in SMF-28e optical fiber, viewed from orthogonal orientations (a and b). (c) illustrates the technique for Gaussian apodization.

The coupling amplitude $\kappa$ of a FBG is proportional to the overlap of the RIM with the intensity profile of the core mode ($\kappa \propto \int \Delta\varepsilon(x,y)|\mathbf{E}(x,y)|^2 \, dx \, dy$, where $\Delta\varepsilon(x,y)$ is the change in the dielectric constant and $\mathbf{E}(x,y)$ is the electric field). Therefore we can control the local coupling amplitude of the grating by varying the transverse offset of the RIMs from the center of the fiber core [8]. This was achieved by applying a voltage to one axis of the piezo-controlled translation stage. We translated the fiber within the focal plane of the objective, as the RIMs have smallest dimensions in this plane due to the ellipsoidal focal volume of a focussed Gaussian beam (see Fig. 1).

The intensity profile of the fiber core mode is approximately Gaussian, and the size of the RIM in the axis of translation (~1 μm) is small with respect to the $1/e^2$ width of the guided mode (10.4 μm). It follows that the coupling strength of the grating varies approximately as a Gaussian function of the offset $x(z)$ of the RIM from the center of the core. Therefore, to inscribe a Gaussian-apodized FBG we simply applied a linear translation of the laser focus across the core during the grating fabrication (see Fig. 1 (c)).

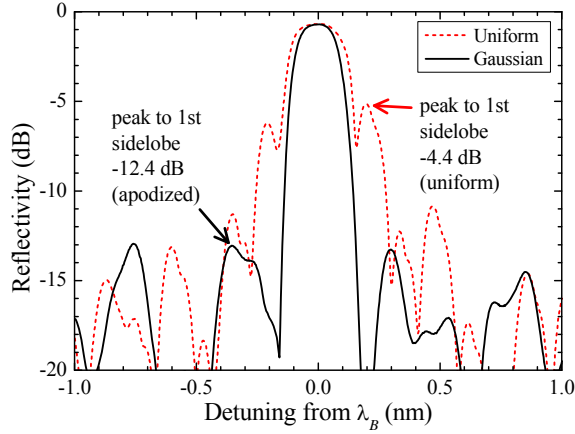

Fig. 2  Reflection spectrum of a Gaussian-apodized FBG and a comparable uniform FBG written with similar inscription parameters.

The measured reflection spectrum of a Gaussian-apodized grating and a comparable uniform FBG is presented in Fig. 2. The Gaussian-apodized FBG had a peak reflectivity of 85%, and the uniform FBG had a peak reflectivity of 86%. The Gaussian-apodized FBG was inscribed with 300 nJ pulses and was 15 mm long with a period of 1.605 μm (third-order resonance at 1549.5 nm). It began and ended with a −7.35 μm and a 7.6 μm offset from the center of the core. The uniform FBG was inscribed with 300 nJ pulses, was 6 mm long and had a period of 1.588 μm (third-order resonance at 1533 nm). The Gaussian-apodized FBG exhibited an 8 dB improvement on side-band suppression compared to the uniform FBG. This is a significant improvement in spectral quality for such a simple fabrication step and could readily improve the performance of fiber laser systems based on PbP FBGs [2,3].

While Gaussian-apodized gratings provide greatly improved side-band suppression over uniform gratings, many applications demand a more tailored spectral response which can only be realized using sophisticated apodization profiles incorporating phase and coupling control. Sinc-apodized gratings are an example of such a grating, exhibiting a characteristic top-hat reflection spectrum which is highly desirable for applications such as band filtering and optical switching.

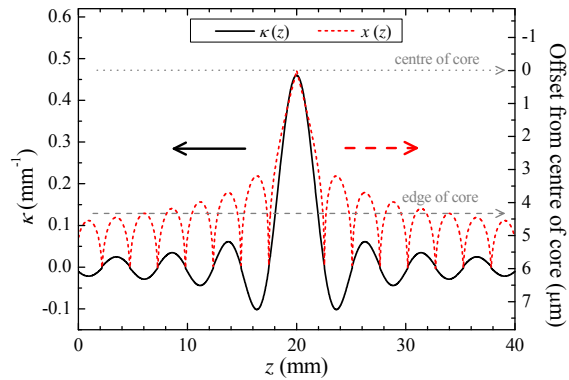

Fig. 3  Coupling amplitude ($\kappa$) profile (black curve) and translation function (red curve) for a sinc-apodized grating.

In order to fabricate FBGs with a sinc apodization profile, it is necessary to introduce $\pi/2$ phase-shifts in the grating to change the sign of the coupling amplitude, as shown in the target $\kappa$ profile in Fig. 3. Thus the desired apodization profile is achieved by combining a translation function $x(z)$ that defines the absolute value of the coupling amplitude, with $\pi/2$ phase-shifts at each of the zero-crossings. We directly control the phase of the grating via the timing of the femtosecond pulse train. To introduce a phase-shift in the grating we switch between two clock signals for triggering the pulses from the femtosecond laser. The two signals both operate at a pulse repetition frequency of 500 Hz, but are offset in time by an amount corresponding to the $\pi/2$ phase-shift.

The desired coupling amplitude profile was $|\kappa(z)| = \kappa_0 |\mathrm{sinc}(2\pi z/(L/N_0))|$ (where $L$ is the length of the grating and $N_0$ is the number of sinc oscillations in the grating). Through the Gaussian form of the guided mode, the actual local coupling amplitude in these PbP gratings is dependent on the offset of the RIMs from the center of the core according to

$$|\kappa(z)| = \kappa_0 \exp\left[-(4x(z)/w)^2\right], \quad (1)$$

where $w$ is the $1/e^2$ width of the core mode (10.4 µm). We thus obtain the translation function

$$x(z) = \frac{w}{4}\sqrt{-\ln|\mathrm{sinc}(2\pi z N_0 / L)|}, \quad (2)$$

but this must be truncated to some maximum value because in the limit as $\kappa \to 0$, $x \to \infty$. However, a truncation of 6 µm or more in this fiber has negligible effect on the grating response. The translation function $x(z)$ and resultant coupling amplitude profile $\kappa(z)$ are shown in Fig. 3.

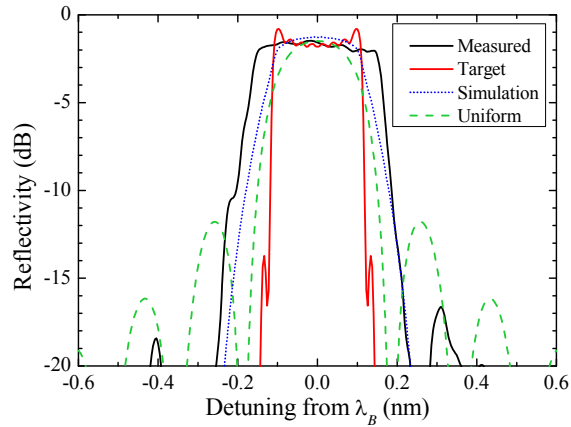

Fig. 4 Measured reflection spectrum of the sinc-apodized grating (solid black), target spectrum based on $\kappa(z)$ shown in Fig. 3 (solid red), simulated spectrum of the fabricated grating (dotted blue) and modeled spectrum of a uniform FBG (dashed green).

We used 220 nJ pulses to inscribe a sinc-apodized grating of length 60 mm and period 1.065 µm (second-order resonance at 1541 nm). Fig. 4 shows the measured reflection spectrum of the grating, the target spectrum (based on the target $\kappa$ profile shown in Fig. 3), a simulation of the fabricated grating, as well as the modeled spectrum of a uniform

FBG with equivalent peak reflectivity and FWHM bandwidth. The modeled spectrum of a uniform FBG highlights the limited spectral characteristics of uniform FBGs, whereas the fabricated grating exhibits all the key qualities of a sinc-apodized FBG: steep band-edges with a relatively wide reflection peak (0.33 nm FWHM) and excellent sidelobe suppression (15.2 dB, peak to first sidelobe). However, by observing the grating with a microscope we found that the amplitude of the translation waveform supplied to the piezo stage was too large, which uniformly stretched the translation function $x(z)$ over a range of 8.4 μm, rather than the intended 6 μm. This resulted in a slightly distorted $\kappa$ profile, hence the discrepancy between the target (solid red curve) and measured spectrum (solid black curve). The simulation of the fabricated grating shown in Fig. 4 (dashed green curve) is based on the measured translation function $x(z)$ of the fabricated grating. This simulation shows good agreement with the measured spectrum, thus we anticipate that future experiments using an appropriate $x(z)$/voltage calibration will yield gratings with steeper band-edges and closer agreement to the design.

In conclusion, we have demonstrated apodized FBGs fabricated using a point-by-point technique. Our method exploits the highly-localized, high-contrast micro-voids formed by focussed IR femtosecond pulse irradiation. We tailor the coupling amplitude by varying the overlap of the micro-void modifications with the core mode and we produce phase-shifts in the grating by introducing discrete time delays to the femtosecond pulse train. We fabricated a sinc-apodized FBG which highlights the strength and the flexibility of our apodization technique, demonstrating that we can implement a custom grating design incorporating phase and coupling amplitude control in order to realize a tailored spectral response.

The ability to inscribe apodized FBGs greatly extends the functionality of the PbP technique. These gratings have immediate application to fiber lasers based on directly-inscribed FBGs for high-power narrow-linewidth operation [2] and all-fiber Q-switched operation [3,12], and indeed any fiber laser or sensing application that requires gratings in non-photosensitive fibers, or gratings with high-temperature stability.

This work was produced with the assistance of the Australian Research Council (ARC) under the Centres of Excellence and LIEF programs, and the German Federal Ministry of Education and Research (BMBF). Christian Voigtländer acknowledges funding by the DAAD (grant D/10/48363) and the Marie Curie foundation within the IRSES project e-FLAG (grant 247635).